\documentclass{iaus}
\usepackage{graphicx}

\title[Lithium depletion signatures in NGC6397] 
{Observational signatures of lithium depletion in the metal-poor globular cluster NGC6397}

\author[Lind et al]   
{Karin Lind$^1$,
      Francesca Primas$^1$, 
      Corinne Charbonnel$^{2,3}$,\\
      Frank Grundahl$^4$,
 \and Martin Asplund$^5$}

\affiliation{$^1$European Southern Observatory, Karl-Schwarzschild-Strasse 2, 
             857 48 Garching bei M\"unchen, Germany  \\ email: {\tt klind@eso.org} \\[\affilskip]
             $^2$Geneva Observatory, 51 chemin des Mailettes, 
             1290 Versoix, Switzerland \\[\affilskip]
             $^3$Laboratoire d'Astrophysique de Toulouse-Tarbes, CNRS UMR 5572, Universit\'e de Tou
louse, 14, Av. E. Belin, F-31400 Toulouse, France \\[\affilskip]
             $^4$Department of Physics \& Astronomy, Aarhus University, Ny Munkegade, 
             8000 Aarhus C, Denmark\\[\affilskip]
             $^5$Max-Planck-Institut f\"ur Astrophysik, Karl-Schwarzschild-Strasse 1,
             857 41 Garching bei M\"unchen, Germany
}

\pubyear{2010}
\volume{268}  
\jname{Light elements in the Universe}
\editors{C. Charbonnel, M. Tosi, F. Primas \& C. Chiappini, eds.}
\begin{document}

\maketitle

\begin{abstract}
The ``stellar'' solution to the cosmological lithium problem proposes that surface depletion of lithium in low-mass, metal-poor stars can reconcile the lower abundances found for Galactic halo stars with the primordial prediction. Globular clusters are ideal environments for studies of the surface evolution of lithium, with large number statistics possible to obtain for main sequence stars as well as giants. We discuss the Li abundances measured for $>$450 stars in the globular cluster NGC\,6397, focusing on the evidence for lithium depletion and especially highlighting how the inferred abundances and interpretations are affected by early cluster self-enrichment and systematic uncertainties in the effective temperature determination.

\end{abstract}

\firstsection 
\section{Introduction}
Through the detailed mapping of the cosmic microwave background performed by the \textit{Wilkinson Microwave Anisotropy Probe} (WMAP), a high-precision estimate of the baryon density of the Universe is nowadays possible. When the most up-to-date prediction, $\Omega_bh^2=0.02273\pm0.00062$ (\cite[Dunkley et al. 2009]{Dunkley09}) is entered in standard Big Bang nucleosynthesis (SBBN), the primordial abundance ratios of D, $^3$He, $^4$He, and $^7$Li with respect to H, are tightly constrained. \cite[Cyburt et al. (2008)]{Cyburt08} thus determine a primordial lithium abundance of $N(\rm^7Li)/\it N(\rm H)=5.24^{+0.71}_{-0.67}\times10^{-10}$ or $A\rm(Li)=2.72\pm0.06$\footnote{$A\rm(Li)=\it\log{\left(\frac{N\rm(Li)}{N(\rm H)}\right)}\rm+12$}, which can be directly compared to the Li abundances inferred for old, metal-poor Galactic halo stars. Numerous high-resolution studies of Li in these stars have indeed revealed a close-to-constant abundance over a wide range in metallicities (forming the ``Spite plateau", originally found by \cite[Spite \& Spite 1982]{Spite82}). However, the observationally inferred values are typically a factor of 3--5 lower than the primordial prediction. Assuming that both the WMAP+SBBN value and the Spite plateau abundances are correct, this implies that the Population II (Pop\,II) stars residing in the halo have undergone significant surface depletion of lithium during their life time. \\

That lithium depletion occurs in solar-type main sequence stars is well-known, e.g. since the solar photospheric value is at least two orders of magnitudes lower than what is found in meteorites. Also in Pop\,I stars that are somewhat hotter than the Sun, characteristic depletion signatures like the famous 'Li dip' are seen in open clusters (first identified in the Hyades by Wallerstein et al. 1965). If indeed lithium depletion takes place also in Pop\,II stars, the process has left little traces behind, since the Spite Plateau is very homogeneous over a wide range in metallicities and effective temperatures, and shows small star-to-star variation. Richard et al.\, (2005) illustrated clearly how atomic diffusion (gravitational settling) of lithium can qualitatively account for the surface drainage, but needs to be moderated by a mixing process below the outer convective envelope to avoid too much Li depletion in stars lying on the hot end of the Spite Plateau. In a series of papers (Talon \& Charbonnel 1998, 2003, 2004), it has been demonstrated that rotation-induced mixing is a plausible source of lithium destruction in both Pop\,I and II stars, given that the rotation of stars cooler than $\sim 6300$\,K also is efficiently stabilised by gravity waves. This stabilising effect is necessary to avoid a strong dependence of the lithium depletion on the initial angular momentum of the stars, which would result in a non-negligible scatter. \\

A next generation of stellar evolution models, accounting self-consistently for atomic diffusion, shear turbulence, and meridional circulation, should be confronted with the detailed lithium abundance patterns found for metal-poor stars. Of fundamental importance for constraining the models are Li abundance trends with effective temperature (which for main sequence stars also is indicative of the stellar mass), with metallicity (which constrains the contribution of post-primordial Li), and with evolutionary phase from the main sequence to the very end of the giant stages of evolution. In Lind et al.\, (2009b) we used FLAMES on VLT/UT2 to investigate the lithium abundances of a very large number of stars in a metal-poor globular cluster (NGC\,6397, [Fe/H]=--2.0), with the primary goals to establish in detail how the abundances depend on evolutionary phase and investigate if there is any significant scatter in abundance for stars in the same phase. In the following we outline our main findings, with focus on the implications and robustness of the observed abundance trends. \\

Fig.\,1 shows how the mean lithium abundance varies with stellar luminosity in NGC6397, from the dwarfs on the main sequence, over the turn-off point, and on the subgiant branch. This figure highlights the need for smooth turbulence, since the pure diffusive models show large gradients that disagree with the observations (models provided by O.\, Richard, \textit{priv. comm}). On the other hand, a non-diffusive model is completely flat, which also is a poor match to the observed trend. A moderate degree of turbulence preserves the flat behaviour on the main sequence, while at least qualitatively reproducing the small upturn located prior to the point of strong depletion, although the location of maximum lithium abundance is not matched. However, the predicted level of depletion is only $\sim$0.2\,dex, which is not enough to reconcile the initial abundance with the cosmological prediction.  

\section{Li deficient un-evolved stars}
Stars in globular clusters are known to show a larger spread in light elements (up to Al) than their counterparts in the halo field. This extra spread is believed to be a result of cluster self-enrichment of the rest products of high-temperature hydrogen burning from an early generation of more massive stars, made possible by the dense intra-cluster environment. Especially N, Na, and Al abundances are enhanced in the most polluted stars, whereas O and Mg levels are depleted. Further, since the stellar ejecta that polluted the cluster gas are presumably void of Li (see e.g. Decressin et al.\, 2007), the Li abundances are theoretically expected to be lower in more polluted stars. Observational evidence of a Li-Na anti-correlation in the globular cluster NGC\,6752 (Pasquini et al.\, 2005) supports this view. \\
In NGC\,6397 we measured for the first time the Li and Na abundances for a large number ($>100$) of turn-off stars and subgiants in a globular cluster. We found indeed, that a handful of rare lithium-deficient stars ($A\rm(Li)<2.0$) also are most strongly enhanced in sodium. However, we also realised that a high degree of pollution is actually necessary to significantly affect the lithium abundances, and that the mean Li abundance trends are essentially unbiased by self-enrichment in NGC6397. We caution however, that this cannot be generalised to other globular clusters, which may have undergone more dramatic enrichment. Lithium abundances in globular clusters should thus preferably be studied in parallel with other light elements.\\
Limiting the sample to stars with $A\rm(Na)<3.9$, having suffered a low degree of pollution, we found that the spread in lithium abundance is very low (0.09\,dex), in accordance with what may be expected from observational uncertainty (see Lind et al.\, 2009b). We therefore expect the real star-to-star scatter in Li abundance in the first generation of cluster stars to be minimal (below 0.05\,dex). \\
Note that both lithium and sodium abundances of metal-poor turn-off stars are sensitive to effective temperature, with $\pm$100\,K corresponding to $\pm$0.07\,dex in Li abundance and $\pm$0.04\,dex in Na abundance. Uncertainties in effective temperature therefore tend to correlate, rather than anti-correlate the abundances. The pollution signature we have identified in NGC6397 is thus robust in this respect. 

\begin{figure}[t]
\begin{center}
 \includegraphics[width=3.0in,angle=90]{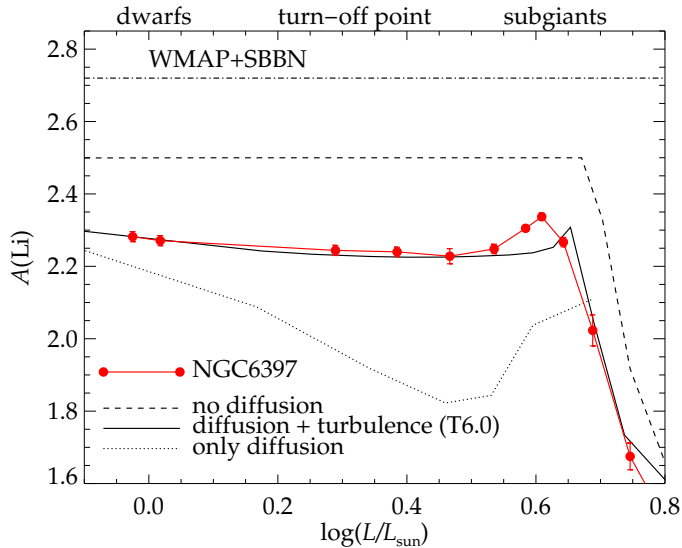}
 \caption{The red bullets represent bin-averaged non-LTE, lithium abundances against stellar luminosity in NGC\,6397 (Lind et al.\, 2009a, 2009b). The dashed line shows the prediction from a model without atomic diffusion (Richard et al.\, 2005), the dotted line corresponds to a model with only diffusion, and the solid line a model with diffusion moderated by turbulence. The initial abundances of the models are shifted by -0.08\,dex to $A\rm(Li)=2.50$. The top line shows the prediction from WMAP+SBBN ($A\rm(Li)=2.72$, Cyburt et al.\, 2008). The difference of 0.22\,dex in Li abundance is still unaccounted for.}
   \label{fig1}
\end{center}
\end{figure}

\section{Li depletion below and above the main sequence turn-off}

\subsection{Subgiants vs turn-off stars}
In canonical models of stellar evolution, assuming convection to be the only mean of particle transport, the surface lithium abundance is constant during the whole main sequence life time, and not until the star has climbed part of the subgiant branch and its convective envelope has expanded to reach Li free layers in the stellar interior, strong surface dilution sets in (the beginning of the ``first dredge up''). This view has been challenged, first by Charbonnel \& Primas (2005) who detected a difference in lithium abundance between subgiant and less evolved stars in the halo field. That subgiant stars are more Li-rich than turn-off stars also in globular clusters was first discovered by Korn et al.\, (2007), for stars in NGC\,6397. In Lind et al.\, (2009b) we used unprecedented number statistics to demonstrate the presence of a gradually increasing trend of Li from the turn-off point to the middle of the subgiant branch, which is there interrupted by the onset of the first dredge-up. These results may indicate that the stars that are subgiants today underwent less efficient Li depletion during their time on the main sequence than the slightly less massive stars now sitting at the turn-off point. An alternative scenario would be that the gravitational settling of Li formed a small over-abundance below the convective envelope which is dredged-up in the subgiants by the penetration of the convective envelope.  \\   
We found that the subgiants are at most 0.1\,dex more Li-rich than the turn-off stars, an estimate which is not suffering from any significant statistical uncertainty, due to the large number of objects ($>200$ subgiants and turn-off stars). However, as is always the case with lithium abundance determination, the systematic uncertainties stemming primarily from the effective temperature determination, are dominant. Since errors in effective temperature and lithium abundance are positively correlated, uncertainties in effective temperatures tend to create artificial trends of higher abundance for stars with higher effective temperature and vice versa. Lithium trends with effective temperature should thus be viewed with great caution. In particular, the true range in $T_{\rm eff}$ over which a change in abundance is discussed must be considerably larger than the uncertainty in relative effective temperature determination. \\
In Lind et al.\, (2009b) we relied on Str\"{o}mgren photometry to infer effective temperatures, and based the final scale on the temperature sensitive $b-y$ index, translated to effective temperatures using synthetic colours computed from MARCS model atmospheres (Gustafsson et al.\, 2008, \"{O}nehag et al.\, 2009). To suppress artificial spread in stellar parameters, caused mainly by photometric uncertainty, we adopted bin-averaged colour-magnitude sequences, constructed from the full photometric catalogue of $\sim5000$ stars. The mean locus of the main sequence turn-off point was thus determined to be 6428\,K ($(b-y)_0=0.303$). In comparison, the new infra-red flux method (IRFM) calibration for dwarfs and subgiants of Casagrande et al.\, (2010), predicts a turn-off point effective temperature of 6435\,K, based on $b-y$. Furthermore, broadband $BVI$ photometry give very similar temperatures when adopting the IRFM calibrations: 6422\,K based on $(B-V)_0=0.386$, and 6447\,K based on $(V-I)_0=0.555$. The photometric evidence thus gives a very consistent view, especially when bearing in mind the different sensitivity to reddening of the colour indices, with $b-y$ the least affected, and $V-I$ the most. \\
The Li abundance difference between subgiants and turn-off stars depends primarily on the temperature difference between the stars. Measuring from the mean locus of the turn-off point ($M_V=4.07$) to the point where the Li abundance reaches its maximum ($M_V=3.4$), the different photometric indices and calibrations listed above predict $\Delta T_{\rm eff}=312\pm40$\,K (mean standard deviation). Taking 40\,K to be an estimate of the uncertainty, the 0.1\,dex difference in $A\rm(Li)$ between turn-off stars and subgiants is significant at the 3$\sigma$ level. \\  
Spectroscopic effective temperatures have been determined by H$\alpha$ profile fitting for stars around the turn-off point in NGC\,6397 in other studies, with resulting scales that are typically $\sim$80\,K cooler than our photometric scale (Korn et al.\, 2007), and $\sim$80\,K hotter (Gonzalez-Hernandez et al.\, 2009). The systematic offsets originate in differences in the adopted model atmosphere and likely also in the quality of the spectroscopic observations. It is important to stress that the study by Gonzalez-Hernandez et al.\, only covered stars above and below but not actually at the turn-off point. The negative slope of A(Li) with decreasing $T_{\rm eff}$ inferred in that study for subgiant branch stars is thus not contradicting the positive increase we identified from the turn-off to the middle of the subgiants branch, contrary to what is claimed by the authors. This is displayed in Fig\, 2, where the left hand panel shows lithium abundance against effective temperature for post turn-off stars in Lind et al.\, and  Gonzalez Hernandez et al.\,, and the right hand panel shows the same abundances, but with absolute visual magnitude as reference scale. The discrepancies seen in the left panel can partly be explained by the fact that their temperature scale is somewhat hotter. Note that the visual magnitudes listed by Gonzalez Hernandez et al. have been adjusted to agree with our absolute scale as described in Lind et al.\, (2009b). \\

\begin{figure}[t]
\begin{center}
 \includegraphics[width=3.0in,angle=90,viewport=0cm 1cm 14cm 25cm]{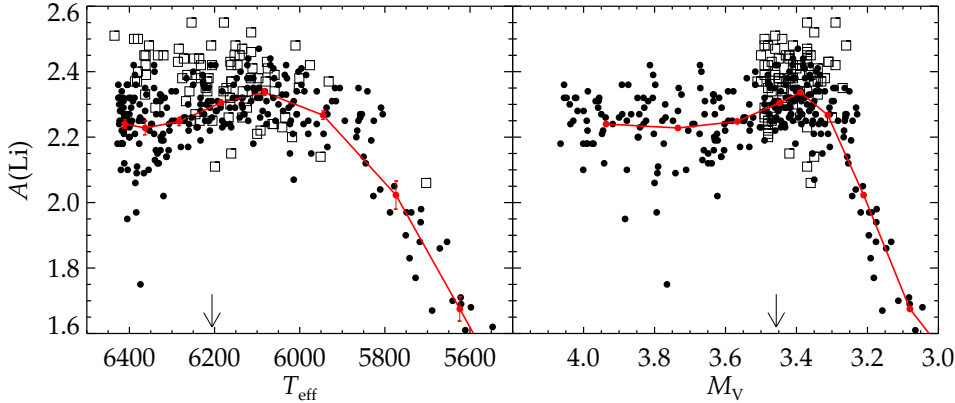}
 \caption{The left hand panel shows lithium abundances vs effective temperature for the post turn-off stars of Lind et al.\, (2009b, black bullets) and Gonzalez-Hernandez et al.\, (2009, open squares). In the right hand panel the same abundances are shown vs visual magnitude. The red, bullets connected with a solid line represent the bin-averaged abundances of Lind et al. The turn-off point is located at $M_V\approx4.07$.}
   \label{fig2}
\end{center}
\end{figure}

\subsection{Hot vs cool dwarf stars}
In Pop\,I dwarfs of a certain age, very lithium-deficient stars are found in the range $T_{\rm eff}=6400-6800$\,K, with a gradual increase on the hot as well as the cool end, shaping the so called Li dip. Talon \& Charbonnel (2004, see also references therein) describe how the shape and location of the dip can be explained when simultaneously accounting for how the efficiencies of convection, rotational-induced mixing, and gravity waves change with effective temperature. On the cool side of the dip, gravity waves dominate the transport of angular momentum, which reduces the differential rotation that otherwise would give rise to large lithium destruction. In Pop\,II stars, an analogous dip is theoretically predicted at the same location, but it has not been detected observationally, because the metal-poor stars that had such hot temperatures on the main sequence have evolved to the giant phases. However, if the hottest metal-poor turn-off stars that exist today (6400-6500\,K) show a lower Li abundance than cooler dwarfs, this would indicate the presence of the dip also in Pop\,II stars. \\
In Lind et al.\, (2009b) we indeed found a 0.04\,dex deficiency in lithium abundance for the hottest dwarfs just below the turn-off at 6430\,K, compared to main sequence stars in the range 6100-6200\,K. However, we cautioned that this difference is very small, and sensitive to errors in the effective temperature scale. In fact, using either of the broad-band indices $B-V$ or $V-I$ to determine effective temperatures enhances the temperature span between the groups enough to erase the difference. The main sequence stars in our sample are thus consistent with having the same lithium abundance.\\

\section{Concluding remarks}
Stellar lithium depletion remains a promising solution to the cosmological lithium problem, but there are several open issues. Non-standard mixing below the convective envelope is obviously of key importance for the surface evolution of lithium in Pop\,II stars. Progress on the modelling side as well as the observational side are needed to establish the underlying physical mechanisms and its correct temperature and metallicity dependence. From the point of view of inferring accurate Li abundances in stars, the effective temperature issue is the most crucial to settle, with the development of hydrodynamical model atmospheres for metal-poor stars being an important step in the right direction.

\end{document}